\documentclass[prl,twocolumn,superscriptaddress]{revtex4-2}
\usepackage{epsfig}
\usepackage{graphicx}
\usepackage{palatino}
\usepackage[english]{babel}
\usepackage{hyphenat}
\usepackage{amsmath}
\usepackage{amssymb}
\usepackage{mathtools}
\usepackage{mathrsfs}
\usepackage{bbm} 
\usepackage{bm}
\usepackage{slashed}
\usepackage{epstopdf}
\usepackage{xcolor}
\usepackage{booktabs}
\usepackage{float}
\definecolor{lcolor}{rgb}{0.,0.0,0.}
\definecolor{citcolor}{rgb}{0,0.,0.5}
\usepackage[breaklinks,colorlinks,urlcolor=blue,citecolor=blue,linkcolor=blue]{hyperref}
\usepackage{multirow}
\usepackage{ltablex}
\usepackage{soul}

\newcommand{\eqn}[1]{Eq.~\eqref{#1}}

\newcommand{\nn}{\nonumber\\ }

\def\be{\begin{eqnarray*}}
\def\ee{\end{eqnarray*}}
\def\beq{\begin{eqnarray}}
\def\eeq{\end{eqnarray}}

\newcommand{\bea}{\beq \begin{aligned}}
\newcommand{\eea}{\end{aligned}\eeq}

\newcommand{\bxi}{{\boldsymbol x_1}}
\newcommand{\bxii}{{\boldsymbol x_2}}

\newcommand{\bx}{{\boldsymbol x}}

\newcommand{\cO}{{\cal O}}

\newcommand{\cU}{{\cal U}}

\newcommand{\rme}{{\rm e}}
\newcommand{\rmd}{{\rm d}}

\newcommand{\rmTr}{{\rm Tr}}

\def\abar{\bar\alpha_s}

\begin{document}

\title{Rotating the Color Glass Condensate} 
\author{Renaud Boussarie}
\email{renaud.boussarie@polytechnique.edu}
\affiliation{CPHT, CNRS, \'Ecole polytechnique, Institut Polytechnique de Paris, 91128 Palaiseau, France}
\author{Paul Caucal}
\email{caucal@subatech.in2p3.fr}
\affiliation{SUBATECH UMR 6457 (IMT Atlantique, Universit\'e de Nantes, IN2P3/CNRS), 4 rue Alfred Kastler, 44307 Nantes, France}
\author{Piotr Korcyl}
\email{piotr.korcyl@uj.edu.pl}
\affiliation{Institute of Theoretical Physics, Jagiellonian University, ul. Lojasiewicza 11, 30-348 Krak\'{o}w, Poland}
\author{Yacine Mehtar-Tani}
\email{mehtartani@bnl.gov}
\affiliation{Physics Department, Brookhaven National Laboratory, Upton, NY 11973, USA}


\begin{abstract}

   High-energy QCD evolution beyond leading order suffers from instabilities driven by large collinear logarithms. We present a framework, consistent with the standard high-energy operator product expansion (OPE), that restores perturbative stability order by order. The method involves a change of basis in the space of high-energy operators, which modifies both the evolution kernel and the coefficient functions while leaving physical observables invariant. Within this factorization scheme, we derive a next-to-leading-order renormalization-group equation whose numerical solution exhibits stable evolution up to large rapidities, paving the way for a systematic framework for precision studies of gluon saturation at current and future colliders.

\end{abstract}

\keywords{Perturbative QCD, factorization, small-x, resummation, BK, BFKL, gluon saturation, CGC}


\maketitle
\section{Introduction}

The partonic picture of hadrons~\cite{Feynman:1969ej,Bjorken:1969ja}, successfully validated in collider experiments---most notably through Deep Inelastic Scattering (DIS) of electrons off protons at high momentum transfer  $Q^2$---is expected to break down at sufficiently high energies. In this regime, \emph{wee} gluons---commonly referred to as small-$x$ gluons, where $x$ denotes the longitudinal momentum fraction of the target proton carried by these gluons---proliferate and their density increases rapidly until non-linear effects associated with gluon recombination become significant and eventually tame the growth of gluon numbers~\cite{Gribov:1984tu,Mueller:1985wy,McLerran:1993ni}. This phenomenon, known as \emph{gluon saturation}, is actively investigated at current facilities~\cite{Morreale:2021pnn} such as RHIC and the LHC, and represents one of the key scientific goals of the future Electron-Ion Collider (EIC)~\cite{AbdulKhalek:2021gbh}. This high-density regime is characterized by the emergence of a new dynamical scale, $Q_s$, the \emph{saturation scale}, whose experimental observation would signal the onset of the saturation phase.

Besides the many experimental challenges involved in probing this emergent regime of QCD dynamics, theoretical progress toward higher-precision calculations has long been hindered by difficulties related to the convergence of perturbation theory in this domain. Within the framework of the Color Glass Condensate (CGC)~\cite{Gelis:2010nm}, the Balitsky--Kovchegov (BK) and Jalilian-Marian--Iancu--McLerran--Weigert--Leonidov--Kovner (JIMWLK) equations~\cite{Balitsky:1995ub,Kovchegov:1999yj,JalilianMarian:1997jx,JalilianMarian:1997gr,Kovner:2000pt,Iancu:2000hn,Iancu:2001ad,Ferreiro:2001qy}, which describe the dense regime of QCD through the high-energy renormalization group (RG) evolution of Wilson-line operators, were found to receive unexpectedly large next-to-leading order (NLO) corrections~\cite{Balitsky:2007feb,Balitsky:2009xg}. These corrections induce severe numerical instabilities~\cite{Lappi:2015fma} and therefore strongly limit the predictive power of the CGC.

This issue was previously diagnosed~\cite{Andersson:1995ju,Kwiecinski:1996td,Kwiecinski:1997ee,Salam:1998tj,Ciafaloni:1998iv,Ciafaloni:1999yw,Ciafaloni:2003rd,SabioVera:2005tiv,Motyka:2009gi} in the linear regime described by the Balitsky--Fadin--Kuraev--Lipatov (BFKL) equation~\cite{Lipatov:1976zz,Kuraev:1977fs,Balitsky:1978ic}, beyond leading order (LO)~\cite{Fadin:1998py,Ciafaloni:1998gs,Fadin:2006ha,Fadin:2007ee,Fadin:2007de}, where it was traced back to large collinear logarithms arising from the potentially wide transverse phase space between the dilute probe---such as a virtual photon with virtuality $Q^2$ in DIS---and the characteristic transverse scale of the target, either of order $\Lambda_{\rm QCD}$  or the nuclear saturation scale $Q_s$.

In the case of a dilute probe with large light-cone positive momentum $q^+$ that scatters off a target carrying large negative light-cone momentum $P^-$, two main pragmatic strategies have been devised to tame these instabilities:  
(i) a modification of the LO BK kernel that effectively resums to all orders the double collinear logarithms  $\alpha_s^n \ln^{2n}(Q^2/Q_s^2)$~\cite{Beuf:2014uia,Iancu:2015vea,Motyka:2009gi}, and  
(ii) a reformulation of the NLO BK equation~\footnote{Also referred to as the next-to-leading logarithmic (NLL) accuracy.} in terms of the target rapidity variable $\ln(P^-/k^-)$  rather than the projectile rapidity variable $\ln(q^+/k^+)$~\cite{Ducloue:2019ezk,Altinoluk:2025tms}.  
While these approaches succeed in stabilizing the NLO evolution~\cite{Ducloue:2019ezk,Lappi:2016fmu,Cepila:2024qge}, they remain unsatisfactory from a formal standpoint.  
On the one hand, the high-energy RG kernel should be universal and uniquely fixed by the structure of the high energy Operator Product Expansion (OPE) \cite{Balitsky:1995ub}; on the other, the OPE for dilute–dense collisions is naturally formulated in projectile-rapidity factorization, enabling a clean separation of coefficient functions at higher orders. Moreover, such modified schemes alter the evolution kernel while ostensibly evolving the same operators as those appearing in the high-energy OPE. Yet these operators arise specifically from $k^+$--space factorization of gluon fields under eikonal kinematics; factorizing in $k^-$--space or adopting modified kinematics would in principle involve a different operator basis. This introduces a fundamental dissonance in the modified frameworks.
Hence, these approaches offer no insight into the all-order structure of high-energy factorization, leaving open the question of their consistency with the underlying OPE. A framework that preserves the structure of the high-energy OPE while maintaining the natural $k^+$--space factorization has therefore remained elusive.

In this Letter, we present a new approach, grounded in the high-energy OPE, that systematically repairs the BK equation order by order in perturbation theory. A detailed derivation and extended discussion are presented in a companion paper~\cite{Boussarie:2025mzh}. It consists of a change of basis in the space of high-energy operators, implemented directly within the OPE.  
Under this transformation, physical cross sections remain invariant, whereas the small-$x$ operators and their corresponding coefficient functions are rotated.  This procedure is analogous to scheme transformations (e.g.~between the $ \text{MS}$  and  $\overline{\text{MS}}$  schemes) familiar from collinear factorization. It can be viewed as the action of an operator ${\rm e}^{-L}$---introduced via the identity $1={\rm e}^{-L} {\rm e}^{L}$---which implements a change of factorization scheme within the high-energy OPE, while leaving physical observables invariant up to higher-order terms in the perturbative expansion.

The operator $L$ effectively reorganizes the perturbative series by introducing a \emph{collinear factorization scale} $\mu^2$, thereby separating transverse (collinear) and longitudinal (rapidity) degrees of freedom in a systematic way. Order by order, $L$ ensures that the rapidity divergences subtracted by the RG evolution are associated with  corresponding collinear counterterms, thus coupling the rapidity scale $\zeta $  to the transverse scale $\mu^2$  through the composite rapidity variable $Y = \ln\left(\zeta/\mu^2\right)$. 
Both the coefficient functions and the evolved operators then acquire a joint dependence on the two scales $(\mu^2, \zeta)$, reflecting the interplay between collinear and small-$x$ logarithmic resummations. Choosing  $\mu^2 = Q^2$ and $\zeta = 2 q^+ P^- = s$ reproduces the standard Bjorken variable, since $\mu^2/\zeta \to Q^2/s =x_{\rm Bj}$, effectively aligning this rapidity subtraction scheme with conventional DIS kinematics.  Yet, we do not propose a new effective theory or a new Wilsonian RG variable. Rather, we construct a new rapidity–collinear factorization scheme within the standard high-energy OPE, defined by an operator-level similarity transformation.

Overall, our framework establishes a unified and transparent formulation of nonlinear QCD evolution in terms of linear-algebraic structures. Within this formulation, specific basis transformations define new factorization schemes that suppress collinear instabilities. The resulting NLO BK equation, derived in this scheme, is shown through detailed numerical simulations to exhibit stable evolution across the relevant physical regime.

\section{Factorization and Scheme Transformation}

Although the framework developed here can be applied to any high-energy observable, we shall illustrate its use by discussing the total DIS cross section in the Regge limit~\cite{Regge:1959mz}, which, after applying the high-energy OPE, takes the following factorized form
 \footnote{An analog  formulation was proposed in the context of jet physics \cite{Becher:2016mmh,Caron-Huot:2015bja}. }
\begin{align}\label{eq:he-factorization}
\sigma(x_{\rm Bj},Q^2) &= \sum_{n=2}^\infty \,\int_{1,...,n } H^{(n)}( Q^2,\{\bx_i\}^n_{i=1},\zeta/s) \nn &\quad\quad \times (S^{(n)}(\{\bx_i\}^n_{i=1},\zeta)-1)\,,
\end{align}
up to power corrections in $x_{\rm Bj}=Q^2/s$, where again $s=2q^+P^-$.
Here $H^{(n)}$ denotes the squared amplitude for a virtual photon with momentum $q\equiv(q^+,-Q^2/2q^+,\boldsymbol{0}_\perp)$ to split into $n$ partons that subsequently interact eikonally with the target, with these interactions encoded in the small-$x$ operators $S^{(n)}$. The latter are built off infinite light-like Wilson lines $U_{\bx}$ ($\cU_{\bx}$) in the fundamental (adjoint) representation at a given transverse coordinate $\bx$ in the background field of the target. The $n$ frozen transverse coordinates of these eikonal partons are gathered in the notation $\{\bx_i\}_{i=1}^n$. For instance, at leading order and next-to-leading order, we have $S^{(2)}(\bx_1,\bx_2) =\frac{1}{N_c}\rmTr \left(U_{\bx_1} U^\dagger_{\bx_2}\right)$ and $S^{(3)}(\bx_1,\bx_2,\bx_3) =\frac{1}{N_c}\rmTr\left(t^a U_{\bx_1} t^b U^\dagger_{\bx_2}\right) \cU^{ab}_{\bx_3}$, respectively.  

Hence, the integer $n$ refers to the number of such Wilson lines in the operator $S^{(n)}$. The convolution is in transverse coordinate space, and we use the shorthand notation $\int_{i}=\int\rmd^2\bx_i/(2\pi)$. Last but not least, $\zeta=2\rho^+P^-$ is the projectile rapidity factorization scale defined in term of the factorization scale $\rho^+$ separating the "fast" ($k^+\ge \rho^+$) and "slow" modes ($k^+\le \rho^+$), the latter being resummed to all orders within the RG evolution equation in $\zeta$ satisfied by $S^{(n)}$. As usual in QCD factorization, $\sigma(x_{\rm Bj},Q^2)$ is $\zeta$-independent, yet the truncation of the perturbative series at a given order, say $\mathcal{O}(\alpha_s^n)$ yields a residual $\zeta$-dependence which is $\mathcal{O}(\alpha_s^{n+1})$. 

We employ the Doi–Peliti formalism~\cite{doi1976,peliti1985}, which uses the Dirac bra–ket notation to describe stochastic dynamics, in order to represent the factorization of the DIS cross section into a hard coefficient function $H$ and a target operator $S$. This will allow one to describe the non-linear QCD dynamics using linear-algebraic techniques and to formulate our change of basis in the space of these operators. 

These objects can be viewed as vectors in a Fock space, denoted respectively by $(H|$ and $|S)$, equipped with the ``dipole-chain'' basis in transverse position space,
\beq
 |0) ,\;|\bx_1\bx_2),\; |\bx_1\bx_2\bx_3),\; |\bx_1\bx_2\bx_3\bx_4) \ldots \,,
\eeq
where now $ \bx_1\bx_2\ldots\bx_n \equiv \{\bx_i\}_{i=1}^n$ denotes an ordered string of transverse coordinates associated with the dipole chain. 
This basis is particularly convenient in the large-$N_c$ limit, where the relevant degrees of freedom are color dipoles in the fundamental representation, i.e.~$S_{12}\equiv S^{(2)}(\bx_1,\bx_2)$. 
For the purely gluonic case, one has the factorized form
\beq\label{eq:dipole-basis}
S^{(n)}\equiv (\bx_1\bx_2\bx_3\ldots\bx_n|S) = S_{12}\,S_{23}\ldots S_{(n-1)n}\,.
\eeq
A fermion loop splits the dipole chain into two segments; higher loops act similarly. For example, a single loop through the shock wave is represented by
\beq
|\bx_1\ldots\bx_i) \otimes |\bx_{i+1}\ldots\bx_n)\,.
\eeq
Within this notation, the cross section can be expressed as an inner product,  $\sigma(x_{\rm Bj},Q^2)=(H(\zeta)|S(\zeta))$. Eq.\,\eqref{eq:he-factorization} then follows by inserting the completeness relation $\mathbb{I}=\sum \int_{1,...,n} |n)(n|$ where the sum runs over all possible (connected and disconnected) dipole chains $|n)$.

At LO, for instance, only $|\bx_1\bx_2)$ contributes to the OPE, with $H^{(2)}=(H|\bx_1\bx_2)$ (see e.g.~\cite{Kovchegov:2012mbw})
and the corresponding operator is $S^{(2)}$:
\bea\label{eq:dipole}
&  S^{(2)}(\zeta)=
(\bx_1\bx_2|S(\zeta)) = \frac{1}{N_c}  \langle \rmTr \, U_{\bxi} U^\dag_{\bxii }\rangle_\zeta\,.
\eea
Here, $\langle \cdots \rangle_{\zeta}$ denotes the renormalized expectation value of the operators at the rapidity scale~$\zeta$ in the target state. The coefficient functions are currently known to NLO accuracy~\cite{Balitsky:2010ze,Beuf:2011xd,Beuf:2016wdz,Beuf:2017bpd,Hanninen:2017ddy,Beuf:2021qqa,Beuf:2022ndu,Beuf:2021srj}, and $n=3$ contributes at this order via the operator $S^{(3)}=S_{13}S_{32}$.

The $\zeta$ dependence of the small-$x$ operators obeys the high-energy RG BK/JIMWLK equation whose solution exponentiates in the space of operators
\begin{align}\label{eq:evolution-op}
|S(\zeta))= & \ \exp\left(\abar  K(\abar) \ln \frac{\zeta}{\zeta_0}\right)\, |S(\zeta_0))\,,
\end{align}
where $K=K_{\rm LO}+\bar\alpha_s K_{\rm NLO}+...$ is the operator generating the Balitsky hierarchy and $\bar\alpha_s=\alpha_s N_c/\pi$.

The action of $K_{\rm LO}$ on the dipole chain basis reads
\begin{align}
  &  (\bx_1\ldots\bx_n|K_{\rm LO}=\sum_{i=1}^{n-1}\int_j\frac{\bx_{i,i+1}^2}{\bx_{ij}^2\bx_{j,i+1}^2}\nn& \, \times\Big[(\bx_1\ldots\bx_i\bx_j\bx_{i+1}\ldots\bx_n| -(\bx_1\ldots\bx_n|\Big]\,,\label{eq:kernel-action}
\end{align}
which for a single dipole yields the LO BK form
\begin{align}\label{eq:K1-action}
    (\bx_1\bx_2|K_{\rm LO}&= \int_3\frac{\bx_{12}^2}{\bx_{13}^2\bx_{32}^2}\, \Big[(\bx_1\bx_3\bx_2| -(\bx_1\bx_2|\Big]\,.
\end{align}
Differentiating Eq.\,\eqref{eq:evolution-op} applied to the dipole w.r.t.~$\ln(\zeta)$ and using \eqn{eq:K1-action} yields the LO BK equation in its conventional form,
\begin{align}
    \frac{\partial S_{12}}{\partial\ln\zeta}&=\abar \int_{3} \frac{\bx_{12}^2}{\bx_{13}^2\bx_{32}^2}(S_{13}S_{32}-S_{12})\,.
\end{align}
We now introduce the ``rotated'' composite operators through the following similarity transformation:
\begin{align}
     |\bar S(\zeta,\mu^2)) =\exp\left(-\bar\alpha_s L(\bar\alpha_s,\mu^2)\right)|S(\zeta))\,,\label{eq:allorder-composite-dipole}
\end{align}
where $L$ is a polynomial in $\ln\mu^2$, with $\mu$ a collinear scale ensuring the consistent matching of rapidity logarithms order by order---effectively redefining the subtraction scheme for rapidity divergences.

In the conventional low-$x$ kinematics, the longitudinal momentum ($k^+$) and the collinear phase space are effectively decorrelated, as reflected in the structure of the evolution equation~\eqn{eq:evolution-op}. While this constitutes a valid factorization scheme, it generates large collinear logarithms at higher orders. In general, the subtraction of logarithmic rapidity divergences is defined only up to an arbitrary logarithms of the transverse degrees of freedom. Physically, 
this transverse log can be linked to constraints on the logarithmic $k^+$ integration---most notably the lifetime condition $k^- = k_\perp^2 / k^+ < P^-$, which implies $k^+ P^- > k_\perp^2$. 
Most {\it ad hoc} schemes effectively impose this kinematic constraint, thereby resumming large collinear logarithms but at the cost of altering the systematics of high-energy factorization and introducing uncontrolled higher-order corrections. 
In contrast, our framework remains fully consistent with high-energy factorization. The operator $L$, whose action is to rotate the dipole-chain basis, is constructed to introduce a counterterm that systematically extracts the collinear logarithm related to the rapidity phase space integral through the lifetime constraint, 
\begin{align}\label{eq:phase-space}
\int_{k_\perp^2/P^-}^{\rho^+}\frac{ \rmd k^+}{k^+}
=\ln \left(\frac{\zeta}{\mu^2}\right)-\ln \left(\frac{k_\perp^2}{\mu^2}\right)\,.
\end{align}
 The first term is associated with the standard rapidity logarithm resummed in the high energy OPE, the second is the collinear logarithm that we aim to reshuffle from the evolution to the hard coefficient function via our scheme transformation $\exp(-L)$. On dimensional grounds, we must therefore introduce a transverse scale $\mu^2$, such that $L\sim \ln k^2_\perp/\mu^2+ \cO(\abar)$ and impose that $\bar S$ be only a function of $\zeta/\mu^2$ beyond leading order. 

To all orders in perturbation theory, we thus build the operator $L=L_{\rm LO}+\abar L_{\rm NLO}+\cO(\abar^2)$ by imposing the composite operators to depend only on the ratio $\zeta / \mu^2$. This requirement entails
\begin{align}
\frac{\partial \bar S}{\partial \ln \mu^2}
+ \frac{\partial \bar S}{\partial \ln \zeta} = 0 \,,
\label{eq:derivative-relation}
\end{align}
which guarantees that any variation of the collinear scale $\mu$ is compensated by a 
corresponding change in the rapidity parameter $\zeta$. 
Eq.\,\eqref{eq:derivative-relation} implies that $\partial L_{\rm LO}/\partial \ln\mu^2=K_{\rm LO}$, such that the constant coefficient of the polynomial in $\ln\mu^2$ is unconstrained reflecting a residual freedom of scheme. To order $\abar$, we shall use 
\begin{align}\label{eq:L1-action}
        &(\bx_1\bx_2|L_{\rm LO} \\&=\int_{3 } \frac{\bx_{12}^2}{\bx_{13}^2\bx_{32}^2}\ln(\mu^2|\bx_{13}||\bx_{32}|) \Big[ (\bx_1\bx_3\bx_2|-(\bx_1\bx_2|\Big]\,,\nonumber
\end{align}
where the argument in the log in Eq.\,\eqref{eq:L1-action} is a scheme choice which turns out to simplify the form of the RG evolution of $\bar S$. \eqn{eq:L1-action} admits a general form similar to Eq.\,\eqref{eq:kernel-action}. 

Expanding the exponential Eq.\,\eqref{eq:allorder-composite-dipole} to first order in $\alpha_s$, we thus obtain $|\bar S)\simeq|S)-\abar L_{\rm LO}|S)$
and therefore
\begin{align}
    &\bar S_{12}\simeq S_{12}\nonumber\\
    &-\abar\int_3 \frac{\bx_{12}^2}{\bx_{13}^2\bx_{32}^2} \ln\left(\mu^2|\bx_{13}||\bx_{32}|\right) \left[ S_{13}S_{32}-S_{12}\right]   \,.\label{eq:composite-dipole-numeric}
\end{align}
 
Evidently, the cross-section must be invariant under the rotation \eqn{eq:allorder-composite-dipole}, which implies the transformation both of the coefficient functions: 
\begin{align}
    (\bar H(\zeta,\mu^2)|=(H(\zeta)|\exp\left[\bar\alpha_s L(\bar\alpha_s,\mu^2)\right]\,,
\end{align}
and of the kernel of the RG equation:
\begin{align}
  \bar K & = \rme^{ - \abar L }  K\,  \rme^{  + \abar L}  \\
  &=  K  + \abar [ K,L]+\frac{\abar^2}{2!}[[K,L],L]+\mathcal{O}(\abar^3)\,,\label{eq:kernel-transformation}
\end{align}
where we have used the Campbell identity. Remarkably, the rotated kernel can be shown to be independent of $\mu$ to all orders in $\abar$, namely $\rmd \bar K / \rmd \ln\mu^2 =0$~\cite{Boussarie:2025mzh}.

We now present the results for the NLO BK equation satisfied by the composite dipole, corresponding to retaining the $\mathcal{O}(\bar\alpha_s)$ terms in Eq.~\eqref{eq:kernel-transformation}. The contribution to $\bar K$ due to the commutator $\bar\alpha_s[K_{\rm LO},L_{\rm LO}]$ can be easily computed using Eq.\,\eqref{eq:kernel-action} for $K_{\rm LO}$ (and $L_{\rm LO}$) (cf. Supplemental Material). Projecting onto the target state $|\bar S)$  and integrating over $\bx_4$ in the two and single dipole contributions \cite{Boussarie:2025mzh}, we obtain our new NLO BK equation 
\begin{widetext} 
\begin{align}
     \frac{\partial \bar S_{12}}{\partial Y}&=\abar  (\bx_1\bx_2|K_{\rm LO}+\abar K_{\rm NLO}|\bar S) + \frac{\abar^2}{2} \int_3\frac{\bx_{12}^2}{\bx_{13}^2\bx_{32}^2} \ln\frac{\bx^2_{13}}{\bx^2_{12}}\ln\frac{\bx^2_{32}}{\bx^2_{12}} (\bar S_{13}\bar S_{32}-\bar S_{12})\,\nn
     +&\frac{\abar^2}{2}\int_{3,4} \frac{\bx_{12}^2}{\bx_{13}^2\bx_{34}^2\bx_{42}^2}\ln\left(\frac{\bx_{34}^2}{\bx_{14}^2}\right)\bar S_{13}(\bar S_{34}\bar S_{42}-\bar S_{32})+\frac{\abar^2}{2}\int_{3,4} \frac{\bx_{12}^2}{\bx_{14}^2\bx_{43}^2\bx_{32}^2}\ln\left(\frac{\bx_{43}^2}{\bx_{42}^2}\right)(\bar S_{14}\bar S_{43}-\bar S_{13})\bar S_{32}+\cO(\alpha_s^3)\,.\label{eq:composite-dipole-RG}
\end{align}
\end{widetext}
The first term represents the full NLO kernel derived in \cite{Balitsky:2007feb}, which contains in $K_{\rm NLO}$ the problematic double logarithms that are exactly canceled by the second term (cf.~Supplemental Material). This cancellation is independent of the running-coupling scheme ultimately used, as long as the same prescription is applied to the $\alpha_s^2$ terms. The last two terms arise from our scheme transformation. Although they do not contain collinear logarithms, they generate anti-collinear double logarithms~\cite{Boussarie:2025mzh} whose numerical effect is discussed in the next section. We further note that $K_{\rm NLO}$ contains additional \textit{single} logarithms~\cite{Iancu:2015joa,Kovner:2023vsy}, which are not canceled by our scheme transformation and whose treatment within our framework is left for future work.  

Our framework unifies two seemingly distinct formulations, each recovered as a limiting case of our construction: \texttt{(i)} the conformal dipole transformation of Ref.~\cite{Balitsky:2007feb}, which introduces an additional factor of $1/2$ in front of the logarithm and in the second term of Eq.~(\ref{eq:derivative-relation})---arising from the scaling $\rho^+/\mu$ instead of $\rho^+ P^-/\mu^2$---together with the scale choice $|\bx_{13}||\bx_{32}|/|\bx_{12}|$; and \texttt{(ii)} the target rapidity evolution of Ref.~\cite{Ducloue:2019ezk} where successive emissions are effectively ordered in $k^-$, corresponding to the scale choice $|\bx_{12}|$ in the logarithm of Eq.~\eqref{eq:composite-dipole-numeric}. For phenomenological applications, these limits must be supplemented by the corresponding transformation $H \to  \bar{H}$ naturally predicted within our formalism.

Finally, in the new basis, the DIS cross section reads
$\sigma(x_{\rm Bj}, Q^2) = \big( \bar{H}(Q^2,\, s/\zeta,\, Q^2/\mu^2) \,\big|\, \bar{S}(\zeta/\mu^2) \big)$, 
where $\bar{H}$ and $\bar{S}$ denote the transformed hard and soft (dipole) operators, respectively. 
To minimize the rapidity logarithms appearing in the hard factor, one may choose $\zeta = s$ and $\mu^2 = Q^2$. 
With this choice, the relation $\mu^2 / \zeta = Q^2 / s = x_{\rm Bj}$ becomes manifest, establishing a direct connection with the Bjorken scaling variable. Hence, our scheme transformation naturally allows both the endpoint of the NLO evolution and the NLO impact factors to be expressed in terms of the natural target variable $x_{\rm Bj}$.

\section{Numerical Stability of NLO BK}

\begin{figure}
    \begin{center}
        \includegraphics[width=0.35\textwidth, angle=270]{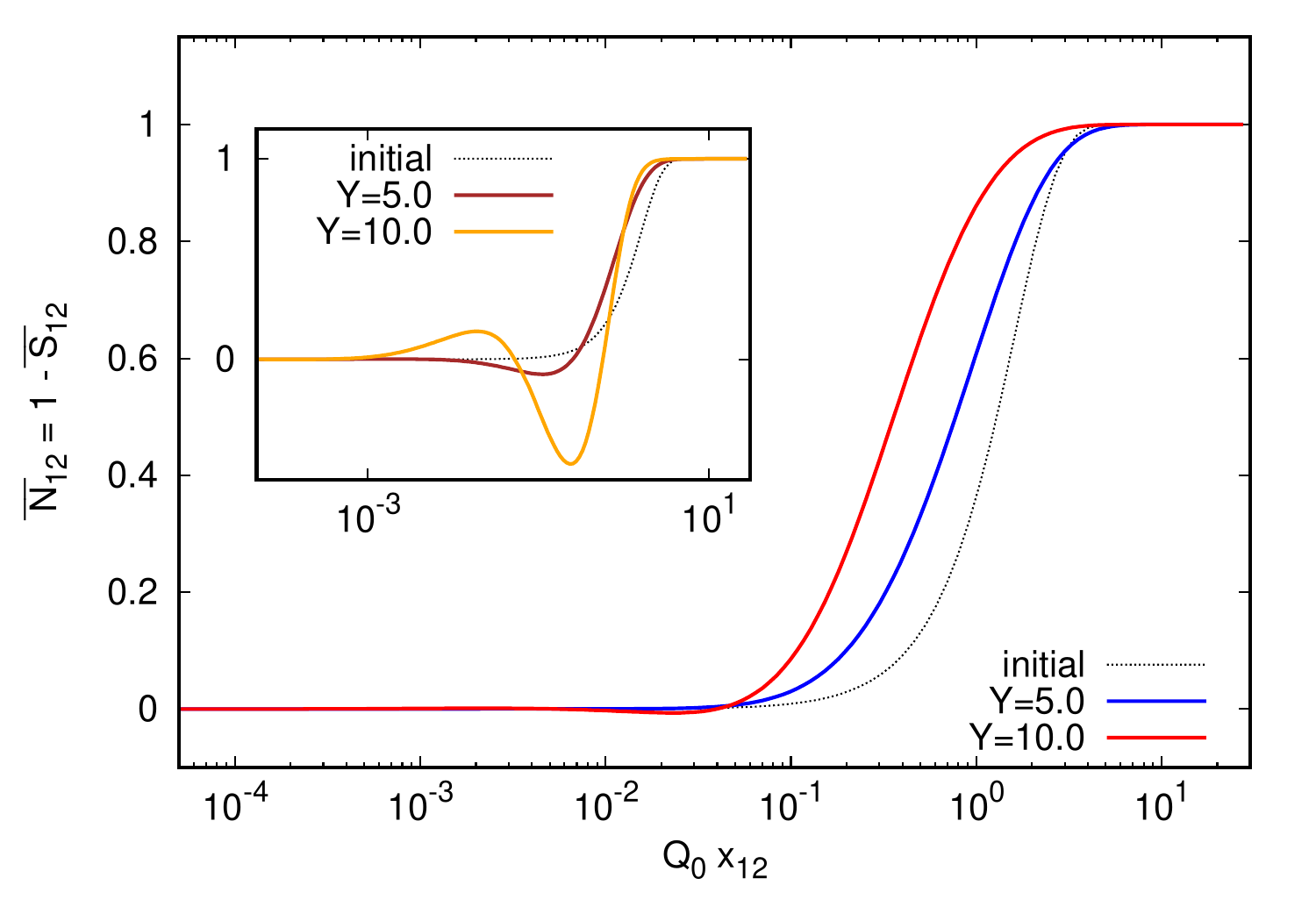}
    \end{center}
    \caption{Evolution of $\bar{N}_{12}$ from the initial condition at $Y=0$ (black dashed line) to $Y=5$ (blue line) and $Y=10$ (red line) using Eq.\,\eqref{eq:composite-dipole-RG}. The abscissa axis is $Q_0x_{12}=Q_0|\bx_{12}|$. 
    The inset shows the evolution from the same initial condition using the standard NLO BK equation. 
    \label{fig:one}}
\end{figure}

We now turn to the numerical analysis of the evolution equation Eq.\,\eqref{eq:composite-dipole-RG}. To demonstrate its stability, we solve it numerically using a simple initial condition inspired by the McLerran–Venugopalan (MV) model \cite{PhysRevD.49.2233,PhysRevD.49.3352},
given by
\begin{equation}
    \bar{S}_{12}(Y=0) = \exp\left[-\frac{1}{4} Q_0^2 \,\bx_{12}^2 \ln\left(\frac{1}{\Lambda |\bx_{12}|} + e\right) \right]\,,
\end{equation}
where $Q_0=1.0$ GeV is the initial saturation scale and $\Lambda=0.2$ GeV is an infrared scale. We employ a fixed coupling constant $\bar{\alpha}_s = 0.3$; the inclusion of running-coupling effects is expected to further enhance the stability of the solution by suppressing short-distance contributions. The convergence of the numerical results with respect to all technical parameters entering the discretized evolution was verified (see the Supplemental Material for a brief discussion of the most relevant tests). The numerical results presented here are intended as a proof of principle; a comprehensive and systematic numerical study of the NLO BK equation, including detailed investigations of initial conditions and running–coupling prescriptions~\cite{PhysRevD.75.014001,IANCU2015643}, is currently underway and will be reported in a forthcoming separate publication. Nevertheless, we have performed preliminary checks showing that these observations are robust under different initial conditions, including the MV and Golec–Biernat–Wüsthoff (GBW) initial conditions \cite{PhysRevD.49.2233,Golec-Biernat:1998zce}, as well as different running-coupling prescriptions, including the Brodsky–Lepage–Mackenzie (BLM) \cite{Brodsky:1982gc} and Balitsky \cite{Balitsky:2006wa} prescriptions. It would be interesting to further solve Eq.~\eqref{eq:composite-dipole-RG} including impact-parameter dependence for a non-homogeneous target, following~\cite{Cepila:2024qge}.

Fig.\,\ref{fig:one} shows the evolved quantity $\bar{N}_{12} = 1 - \bar{S}_{12}$ for two representative rapidities, $Y = 5.0$ and $Y = 10.0$. The evolution remains stable throughout the phenomenologically relevant rapidity range, extending beyond the kinematic reach of existing DIS measurements at HERA~\cite{H1:2015ubc}, up to $Y = \ln(\zeta / \mu^2) \simeq 10$. At larger rapidities, residual anti-collinear double-logarithmic and collinear single logarithmic instabilities begin to appear, signaling the onset of higher-order effects. Various strategies have been proposed to address these anti-collinear logarithms through an all-order resummation implemented in the kernel of the RG equation~\cite{Iancu:2015vea,Ducloue:2019ezk}. This breakdown of truncated perturbation theory is, in any case, expected: as $Y$ increases—typically for $Y \gtrsim 1/\alpha_s^3$—next-to-next-to-leading order (NNLO) and higher-order corrections become important and must be included. 

The inset in Fig.\,\ref{fig:one} displays the results obtained using the same initial condition evolved with the standard NLO BK equation. In this case, as first noticed in~\cite{Lappi:2015fma}, the double-logarithmic contribution dominates the evolution and rapidly drives $\bar{N}_{12}$ to unphysical negative values.

\section{Conclusions}

In summary, we have proposed a systematically improvable framework that restores the perturbative stability of high-energy QCD evolution at next-to-leading order and beyond, while remaining consistent with the high-energy OPE. The key ingredient is a rotation in the space of high-energy operators, implemented by the operator $e^{-L}$, which preserves physical cross sections while reorganizing the factorization such that the large collinear logarithms responsible for the instabilities of the conventional NLO BK evolution are eliminated. Applied to the dipole operator, this transformation yields an evolution equation that is both stable and positive across the phenomenologically relevant rapidity range, as confirmed by dedicated numerical simulations.

Beyond the specific NLO application, our formalism recasts nonlinear small-$x$ dynamics into the language of linear algebra: operators are represented in Dirac notation and act on a vector space spanned by dipole-chain states, providing a compact and transparent framework for describing their evolution and interactions. Within this operator-based picture, the change of basis underlying the factorization scheme transformation becomes explicit, offering a unified algebraic interpretation of various high-energy evolution schemes. 

This approach provides a blueprint for systematically extending perturbative stability to N$^2$LO and beyond, and for consistently incorporating single-logarithmic and running-coupling effects. In particular, together with the recent computation of the BK kernel at $\mathrm{N}^2$LO accuracy \cite{Caron-Huot:2016tzz,Brunello:2025rhh}, our method opens the way for the first stable numerical investigations of the evolved dipole at this precision. These developments represent an important step toward establishing the theoretical consistency and numerical reliability required for future precision studies of gluon saturation and small-$x$ dynamics at current and future facilities, notably the EIC, and can be readily extended to other observables such as inclusive dijet, forward hadron production or semi-inclusive DIS.

\vspace{0.5cm}
\begin{acknowledgments}
\noindent{\bf Acknowledgements.} We thank Guillaume Beuf, Giovanni Chirilli, Edmond Iancu,  Farid Salazar, Anna Sta\'sto, Leszek Motyka, and Dionysis Triantafyllopoulous for valuable discussions related to this work. P.~C. is funded by the Agence Nationale de la Recherche under
grant ANR-25-CE31-5230 (TMD-SAT). Y.~M.~T. was supported by the U.S. Department of Energy under Contract No. DE-SC0012704 and by Laboratory Directed Research and Development (LDRD) funds from Brookhaven Science Associates. We are grateful for the support of the Saturated Glue (SURGE) Topical Theory Collaboration, funded by the U.S. Department of Energy, Office of Science, Office of Nuclear Physics. Numerical calculations were performed on the LUMI supercomputer under the time allocation: project\_465002091 (Calculating predictions for EIC physics). We gratefully acknowledge Polish high-performance computing infrastructure PLGrid (HPC Center: ACK Cyfronet AGH) for providing computer facilities and support within the computational grant no. PLG/2024/017690. P.~K. was supported by the Polish National Science Center (NCN) grant No. 2022/46/E/ST2/00346. P.~C. and P.~K. thank the EIC theory institute at BNL for its support and hospitality.
\end{acknowledgments}

\bibliographystyle{apsrev4-1}

\bibliography{nlobk-references.bib}

\newpage

\phantom{end page}

\newpage

 \section{Supplemental material}
 \subsection{Numerical setup}

 For completeness, we describe here the numerical procedure used in our simulations in more detail. The transverse area integrals on the right-hand side of the evolution equation Eq.\,\eqref{eq:composite-dipole-RG} are evaluated using polar coordinates. The radial variables are discretized on a logarithmic grid. The solution shown in Fig.~\ref{fig:one} was obtained using a grid of 256 nodes for $|\bx_3|$ and $|\bx_4|$. We limit these variables to the interval $[10^{-7},30]$. The angular degrees of freedom $\phi_3$ and $\phi_4$ were discretized using 16 nodes. We employ the Simpson $\frac{3}{8}$ quadrature scheme for both the angular and the radial variables, i.e., each angular variable is sampled at 48 points.

 The choice of the coordinate systems for $\bx_3$ and $\bx_4$ is made such that the coordinates $\bx_1$ and $\bx_2$ lie along the x-axis, with the origin set at $\frac{1}{2}(\bx_1+\bx_2)$. In the two integrals involved in the $K_{1234}$ and $K_f$ terms, the position $\bx_4$ is measured with respect to the position $\bx_3$. The latter choice is dictated by the fact that in such a coordinate system, one can analytically cancel the divergence $1/\bx_{34}^4$ in the limit of small $\bx_{12}$, which otherwise becomes numerically unstable.

 \begin{figure}
     \begin{center}
         \includegraphics[width=0.35\textwidth, angle=270]{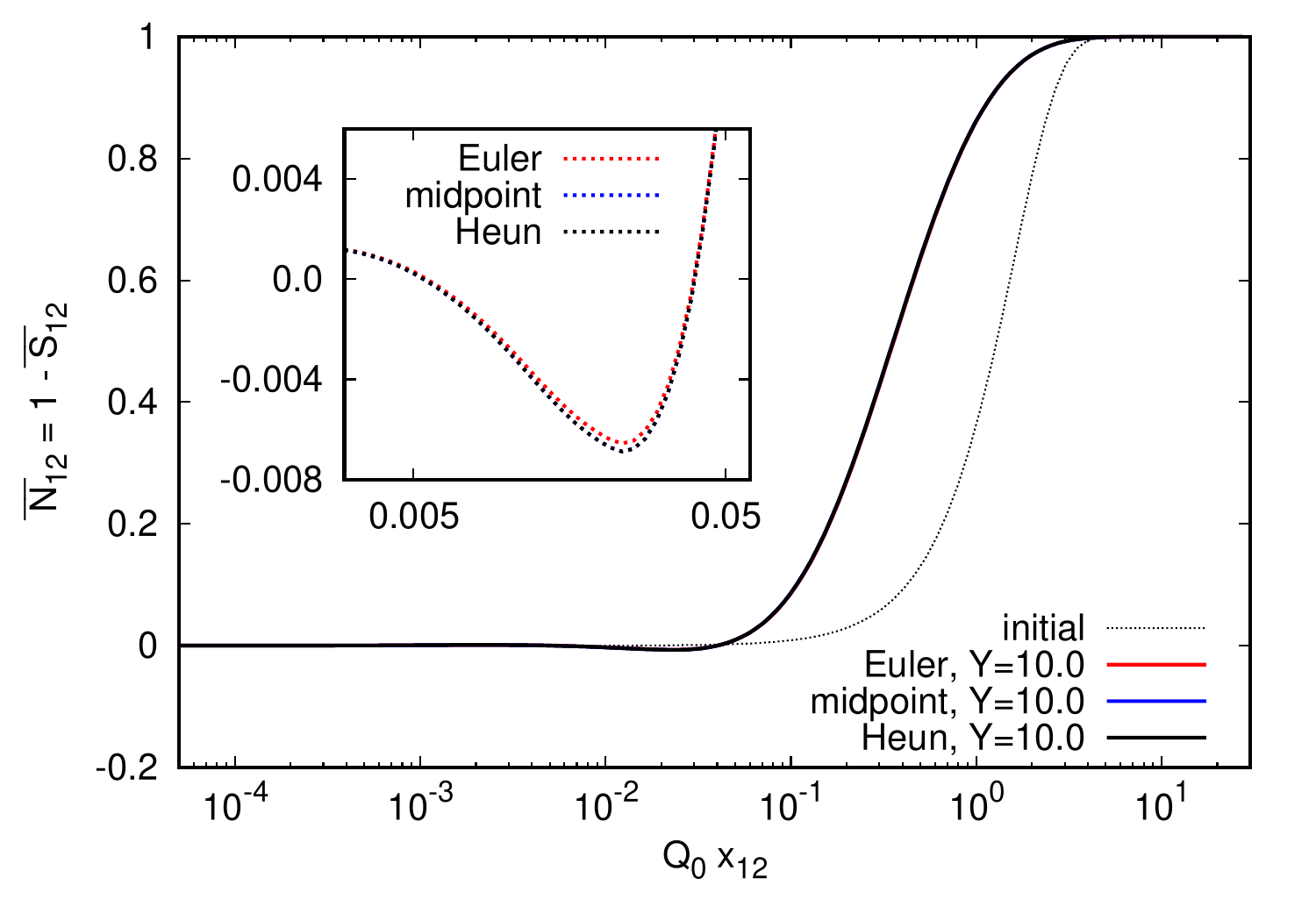}
     \end{center}
     \caption{The evolution of $\bar{N}_{12}$ from the initial condition up to $Y=10.0$ with $N=256$. Black dashed curve denotes the initial condition, while the red, blue, and black data correspond to the Euler, midpoint, and Heun methods, respectively.  The difference between the Euler method and the other two is largest in the small region around $Q_0 x_{12} \approx 0.02$, but remains negligible overall.
     \label{fig:two}}
 \end{figure}

 \begin{figure}
     \begin{center}
         \includegraphics[width=0.35\textwidth, angle=270]{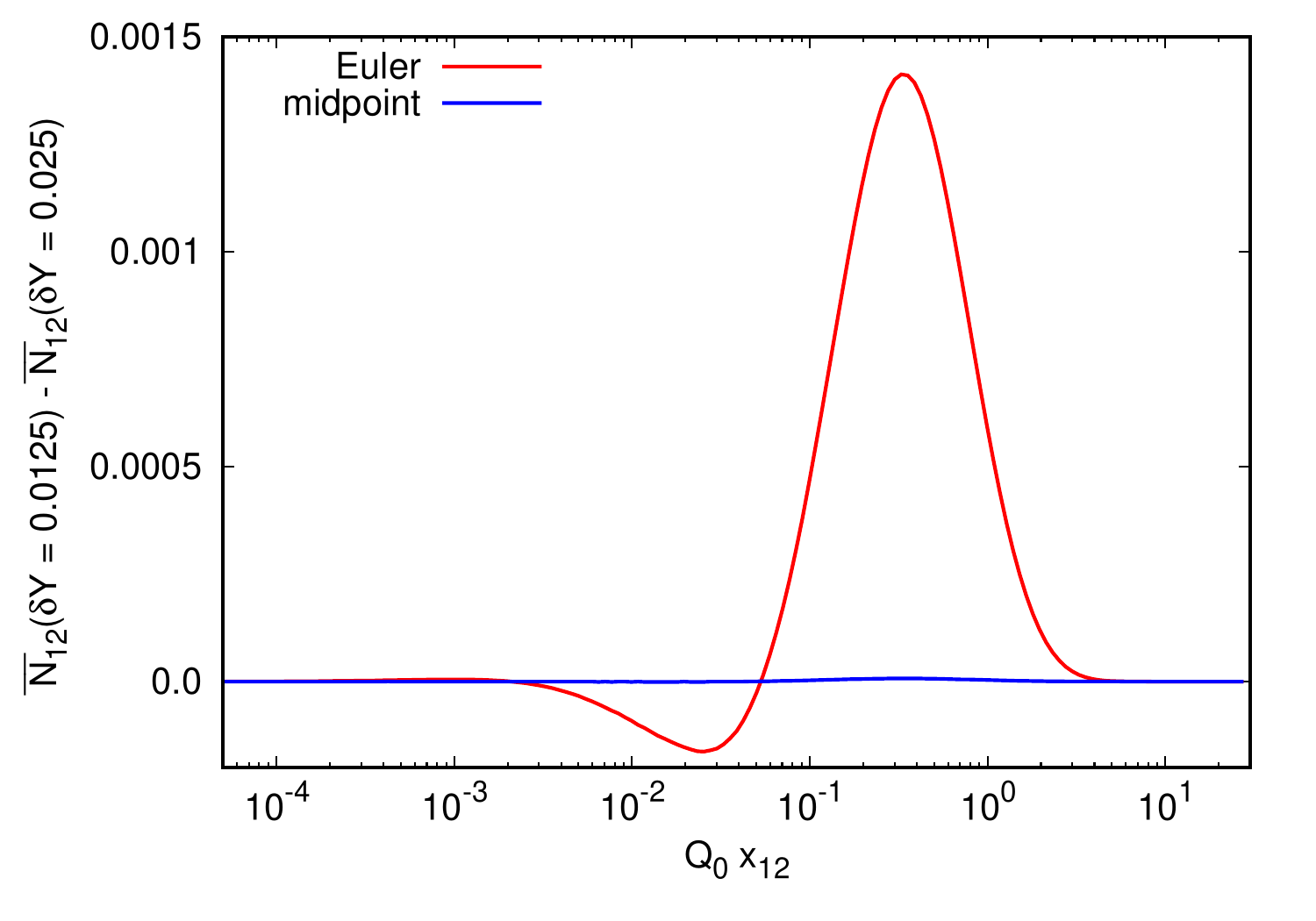}
     \end{center}
     \caption{The evolution of $\bar{N}_{12}$ from the initial condition up to $Y=10.0$ with $N=256$. Plotted is the difference between the results obtained with $\delta Y=0.025$ and $\delta Y=0.0125$. As expected, the convergence of the Euler method is slower compared to the other methods. The maximal difference for the midpoint method is of the order of $10^{-5}$. 
     \label{fig:three}}
 \end{figure}

 The differential equation is integrated numerically, i.e. the rapidity variable $Y$ is discretized with some small step $\delta Y$, $Y_n = Y_0 + n \delta Y$, and the amplitude $S_{12}(Y_{n+1}=Y_n+\delta Y)$ is obtained from its previous state $S_{12}(Y_n)$ using one of the three integration rules: Euler, Heun, or midpoint. The Euler method advances the solution through 
 \begin{equation*}
 S_{12}(Y_{n+1}) = S_{12}(Y_n) + \delta Y f(S_{12}(Y_n)),
 \end{equation*}
 the Heun method through 
 \begin{multline*}
 S_{12}(Y_{n+1}) = S_{12}(Y_n) +\\+ \frac{\delta Y}{2} \big\{ f(S_{12}(Y_n)) + f\big[ S_{12}(Y_n) + \delta Y f(S_{12}(Y_{n})) \big] \big\},
 \end{multline*}
 whereas the midpoint method uses 
 \begin{equation*}
 S_{12}(Y_{n+1}) = S_{12}(Y_n) + \delta Y f\big[ S_{12}(Y_n) + \frac{\delta Y}{2} f(S_{12}(Y_{n}))\big].
 \end{equation*}
 The increment step $\delta Y$ was chosen such that all three methods yield equivalent solutions. In Fig.\,\ref{fig:two} we compare the amplitude at $Y=10.0$ obtained with the three methods. The largest deviation can be seen between the solution obtained with the Euler method and the solutions obtained using the midpoint and Heun methods. In Fig.\,\ref{fig:three}, we show the difference between the resulting $N_{12}(Q_0 x_{12}, Y=10.0)$ evolved with the step $\delta Y = 0.0125$ or $\delta Y = 0.025$. As expected, the midpoint method exhibits a difference of the order of $10^{-5}$ because of its faster convergence.

 \begin{figure}
     \begin{center}
         \includegraphics[width=0.35\textwidth, angle=270]{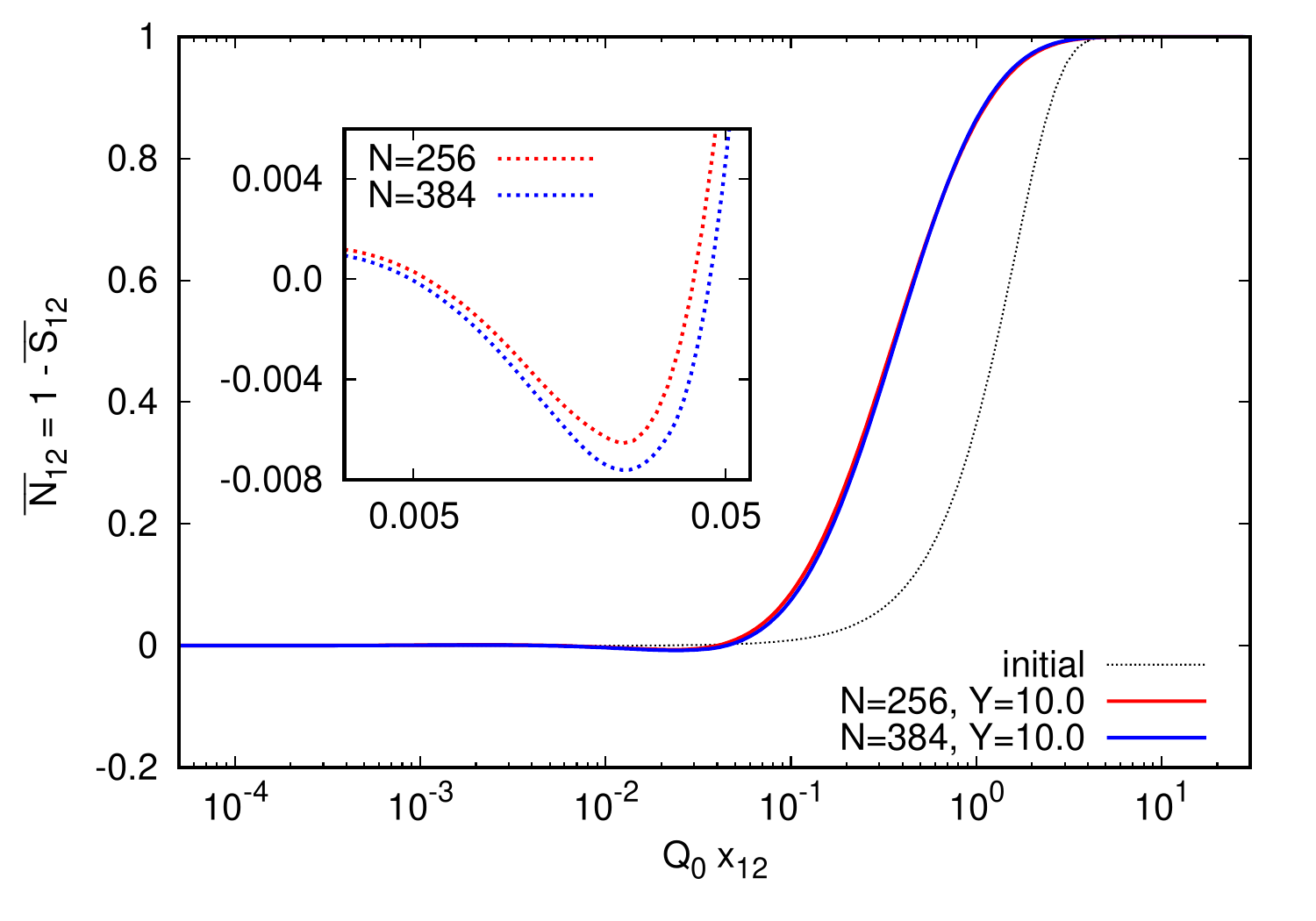}
     \end{center}
     \caption{The evolution of $\bar{N}_{12}$ from the initial condition up to $Y=10.0$ using the Euler method and two different grid sizes: $N=256$ (red data) and $N=384$ (blue data). The largest deviation of the order of $0.002$ is observed around $Q_0 x_{12} \approx 0.1$, but again remains negligible overall. 
     \label{fig:four}}
 \end{figure}

 We checked that increasing the resolution in the discretization of all degrees of freedom and extending the interval for the radial variables does not alter the resulting dipole. Eventually, we also tested that running the evolution in double and long double precision yields compatible results. An example of such tests is shown in Fig.\,\ref{fig:four}, where we compare the resulting amplitude obtained with two different grid sizes.\\

 \subsection{NLO BK kernel in the dipole chain notations}\label{app:NLO-BK}

 In the Doi-Peliti formalism, the NLO correction $K_{\rm NLO}$ of the BK kernel reads (setting $\beta_0=0$ for fixed coupling):
 \begin{eqnarray}
 (\boldsymbol{x}_{1}\boldsymbol{x}_{2}|K_{\rm NLO} & = &\frac{1}{2} \int_{3}K_{123}^{\rm NLO}\left[(\boldsymbol{x}_{1}\boldsymbol{x}_{3}\boldsymbol{x}_{2}|-(\boldsymbol{x}_{1}\boldsymbol{x}_{2}|\right]\label{eq:BKkernelNLL}\\
  & + & \frac{1}{2}\int_{3,4}K_{1234}\left[(\boldsymbol{x}_{1}\boldsymbol{x}_{3}\boldsymbol{x}_{4}\boldsymbol{x}_{2}|-(\boldsymbol{x}_{1}\boldsymbol{x}_{3}\boldsymbol{x}_{2}|\right]\nonumber \\
  & + & \frac{n_{f}}{2N_{c}}\int_{3,4}K_{f}(\boldsymbol{x}_{3}\boldsymbol{x}_{2}|\otimes\left[(\boldsymbol{x}_{1}\boldsymbol{x}_{4}|-(\boldsymbol{x}_{1}\boldsymbol{x}_{3}|\right]\,,\nonumber 
 \end{eqnarray}
 with
 \begin{align}
 &K_{123}^{\rm NLO}=\frac{\bx_{12}^2}{\bx_{13}^2 \, \bx_{32}^2} \left[
  \frac{67}{18} - \frac{\pi^2}{6} - \frac{5}{9} \frac{n_f}{N_c} 
 -\ln \frac{\bx_{13}^2}{\bx_{12}^2} \ln \frac{\bx_{32}^2}{\bx_{12}^2} 
 \right]\,,\label{eq:K123-NLO-def}\\
     &K_{1234}= - \frac{2}{\bx_{34}^4} 
 + \left[ 
 \frac{ \bx_{13}^2 \bx_{24}^2 + \bx_{14}^2 \bx_{32}^2 
 - 4 \bx_{12}^2 \bx_{34}^2 }
 { \bx_{34}^4 (\bx_{13}^2 \bx_{24}^2 - \bx_{14}^2 \bx_{32}^2) } \right.\nonumber\\
 &\left.+ \frac{ \bx_{12}^4 }{ \bx_{13}^2 \bx_{24}^2 (\bx_{13}^2 \bx_{24}^2 - \bx_{14}^2 \bx_{32}^2) } 
 + \frac{ \bx_{12}^2 }{ \bx_{13}^2 \bx_{24}^2 \bx_{34}^2 } 
 \right]\ln \frac{ \bx_{13}^2 \bx_{24}^2 }{ \bx_{14}^2 \bx_{32}^2 }\,,\\
 &K_f = \frac{2}{\bx_{34}^4} 
 - \frac{ \bx_{14}^2 \bx_{32}^2 + \bx_{24}^2 \bx_{13}^2 - \bx_{12}^2 \bx_{34}^2 }
 { \bx_{34}^4 (\bx_{13}^2 \bx_{24}^2 - \bx_{14}^2 \bx_{32}^2) } 
 \ln \frac{ \bx_{13}^2 \bx_{24}^2 }{ \bx_{14}^2 \bx_{32}^2 }\,.
 \end{align}
 Eq.\,\eqref{eq:BKkernelNLL} can be generalized to any $(\bx_1 \bx_2 \ldots \bx_n |$ state by following the same underlying Leibniz rule as in Eq.\,\eqref{eq:kernel-action}.

 \subsection{Computation of $[K_{\rm LO},L_{\rm LO}]$}\label{app:commutator}
 We briefly derive the commutator $[K_{\rm LO},L_{\rm LO}]$ in the dipole-chain notation to illustrate the method. For convenience, we further simplify the notation by introducing the shorthand
 \begin{align} 
 &(\bx_1\bx_2| \equiv (12|\,, \quad (\bx_1\bx_3\bx_2| \equiv (132|\,, 
 \end{align}
 and similarly for higher dipole states.
 We further write
 \begin{align}
 K_{123} =  \frac{\bx^2_{12}}{\bx^2_{13}\bx^2_{32}}\,,
 \end{align}
 and 
 \begin{align}
 L_{123} =  \frac{\bx^2_{12}}{\bx^2_{13}\bx^2_{32}} \ln\left(\mu^2 |\bx_{13}||\bx_{32}|\right)\,.
 \end{align}
 Using \eqn{eq:K1-action} and \eqn{eq:L1-action} we readily obtain 

 \begin{align}
     (12|K_{\rm LO}L_{\rm LO}=\int_{3,4}
     & \Big\{K_{123}L_{134}\big[(1432| -(132|\big]\nn &
     +K_{123}L_{324}\big[(1342|-(132|\big]\nn&
     -K_{123} L_{124} \big[(142| -(12|\big] \Big\}\,,
 \end{align}
 and
 \begin{align}
     (12|L_{\rm LO}K_{\rm LO}=\int_{3,4}
     & \Big\{L_{123}K_{134}\big[(1432| -(132|\big]\nn &
     +L_{123}K_{324}\big[(1342|-(132|\big]\nn&
     -L_{123} K_{124} \big[(142| -(12|\big]\,,
 \end{align}
 such that the commutator is
 \begin{align}
    & (12|[K_{\rm LO},L_{\rm LO}]\nn
     &=\int_{3,4}
      \Big\{(K_{123}L_{134}-L_{123}K_{134})\big[(1432| -(132|\big]\nn &
     +(K_{123}L_{324}-L_{123}K_{324})\big[(1342|-(132|\big] \nonumber\\
     &-(K_{123}L_{124}-L_{123}K_{124})\big[(142|-(12| \big]\Big\}\\
     &=\int_{3,4} \Big\{ \frac{\bx_{12}^2}{\bx_{14}^2\bx_{43}^2\bx_{32}^2} \ln\frac{|\bx_{14}||\bx_{43}|}{|\bx_{13}||\bx_{32}|}
     \big[(1432| -(132|\big]\nn &
     + \frac{\bx_{12}^2}{\bx_{13}^2\bx_{34}^2\bx_{42}^2} \ln\frac{|\bx_{34}||\bx_{42}|}{|\bx_{13}||\bx_{32}|}\big[(1342|-(132|\big] \nonumber\\
     &-\frac{\bx_{12}^4}{\bx_{13}^2\bx_{32}^2\bx_{14}^2\bx_{42}^2}\ln\frac{|\bx_{14}||\bx_{42}|}{|\bx_{13}||\bx_{32}|}\big[(142|-(12|\big]\Big\}\,.
 \end{align}
 Once contracted with the state $|\bar S)$, this equation yields Eq.\,(111) of~\cite{Boussarie:2025mzh}, and eventually, Eq.\,\eqref{eq:composite-dipole-RG} of this paper (see~\cite{Boussarie:2025mzh} for more details).
\end{document}